\documentclass[prl,twocolumn,showpacs,superscriptaddress,floatfix]{revtex4-1}
\usepackage{graphicx,amsfonts,amssymb,amsmath}
\usepackage[colorlinks=true,citecolor=green,linkcolor=red,urlcolor=blue]{hyperref}

\newcommand{\ba}{{\boldsymbol{a}}}
\newcommand{\bb}{{\boldsymbol{b}}}

\newcommand{\bk}{{\boldsymbol{k}}}
\newcommand{\bK}{{\boldsymbol{K}}}

\begin{document}

\title{Fate of Dirac points in a vortex superlattice}

\author{Michael Kamfor}
\email{kamfor@fkt.physik.tu-dortmund.de}
\affiliation{Lehrstuhl f\"{u}r Theoretische Physik I, Otto-Hahn-Stra\ss e 4, TU Dortmund, 44221 Dortmund, Germany}
\affiliation{Laboratoire de Physique Th\'eorique de la Mati\`ere Condens\'ee,
CNRS UMR 7600, Universit\'e Pierre et Marie Curie, 4 Place Jussieu, 75252 Paris Cedex 05, France}

\author{S\'ebastien Dusuel}
\email{sdusuel@gmail.com}
\affiliation{Lyc\'ee Saint-Louis, 44 Boulevard Saint-Michel, 75006 Paris, France}

\author{Kai Phillip Schmidt}
\email{schmidt@fkt.physik.tu-dortmund.de}
\affiliation{Lehrstuhl f\"{u}r Theoretische Physik I, Otto-Hahn-Stra\ss e 4, TU Dortmund, 44221 Dortmund, Germany}

\author{Julien Vidal}
\email{vidal@lptmc.jussieu.fr}
\affiliation{Laboratoire de Physique Th\'eorique de la Mati\`ere Condens\'ee,
CNRS UMR 7600, Universit\'e Pierre et Marie Curie, 4 Place Jussieu, 75252 Paris Cedex 05, France}


\begin{abstract}
We consider noninteracting fermions on the honeycomb lattice in the presence of a magnetic vortex superlattice. It is shown that depending on the superlattice periodicity, a gap may open at zero energy. We derive an expression of the gap in the small-flux limit but the main qualitative features are found to be valid for arbitrary fluxes.
This study provides an original example of a metal-insulator transition induced by a strongly modulated magnetic field in graphene. At the same time our results directly apply to Kitaev's honeycomb model in a vortex superlattice.
\end{abstract}

\pacs{73.22.Pr, 71.20.-b,71.10.-w,}

\maketitle



After its experimental discovery by Geim and Novoselov \cite{Novoselov04} in 2004, graphene's electronic properties received much attention (see Ref.~\cite{CastroNeto09} for a review). 
However, the band structure of this honeycomb lattice in the tight-binding approximation has been known for several decades, \cite{Wallace47} and its modifications in the presence of a uniform magnetic field were investigated more than 20 years ago by Rammal \cite{Rammal85}. 
One of the most salient features of the zero-field spectrum is the existence of a point-like Fermi surface at zero energy, the celebrated Dirac points, giving rise to a relativistic dispersion in their neighborhood (the so-called Dirac cones). Interestingly, these discrete zero-energy states are still present when a uniform magnetic field, is added \cite{Rammal85,Esaki09}. The stability of these states has led several groups to analyze the influence of a nonuniform magnetic field and it is now commonly accepted that a {\em smoothly} modulated magnetic field is not sufficient to open a gap at zero energy \cite{Snyman09,Guinea10,Lin11}. 

\begin{figure}[t]
\includegraphics[width=0.8\columnwidth]{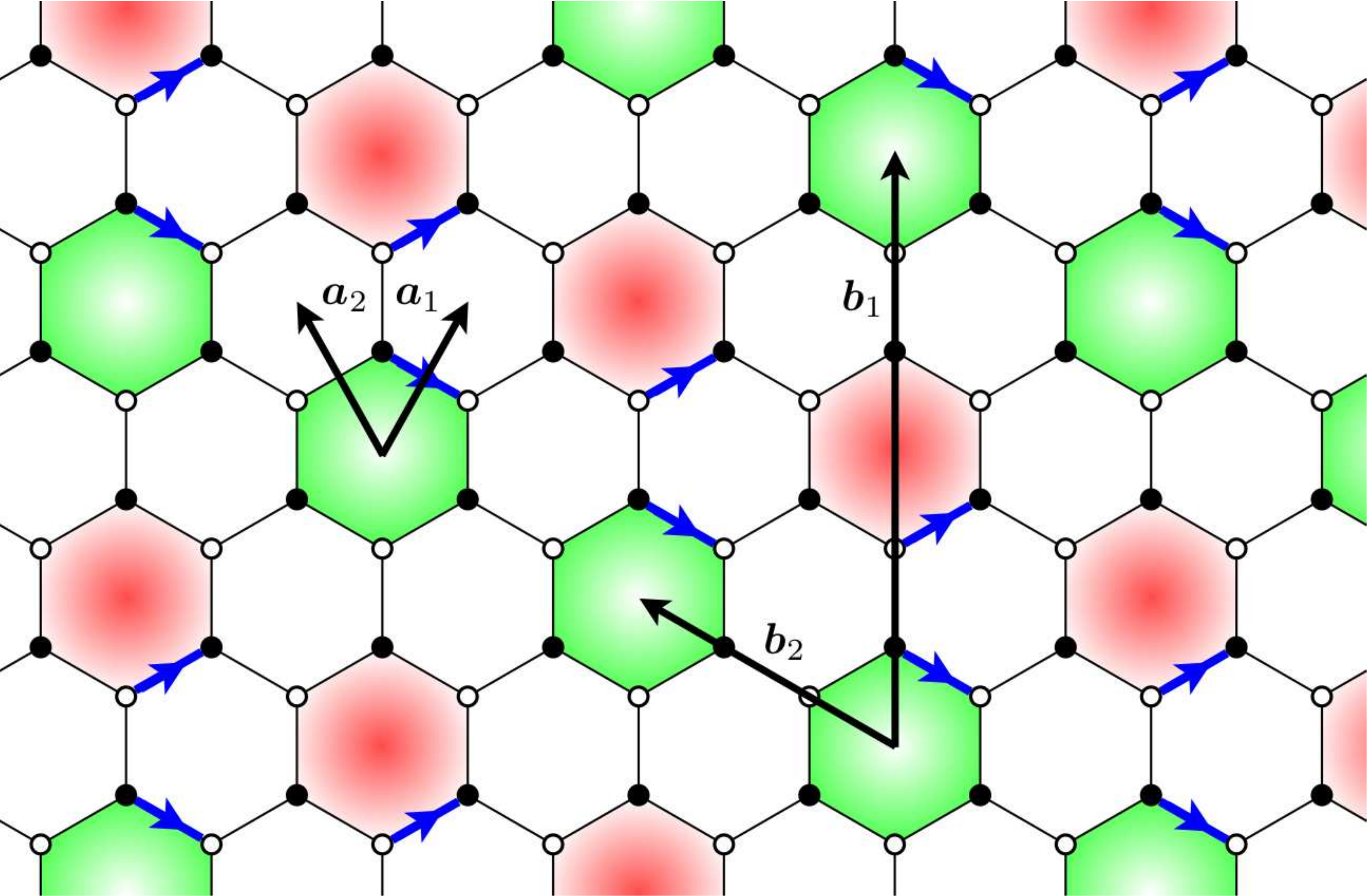}
\caption{(Color online) A piece of the \mbox{${\mathcal L}(p=1,q=1)$} magnetic vortex superlattice  spanned by primitive vectors $\bb_1$ and $\bb_2$. Vectors $\ba_1$ and $\ba_2$ are primitive vectors of the bare honeycomb lattice. Light-center (green) and dark-center (red) plaquettes contain a flux $+\phi$ and $-\phi$, respectively, whereas white plaquettes are flux-free. 
Blue links with arrows indicate oriented hopping terms ``carrying" the flux.}
\label{fig:lattice}
\end{figure}

In this paper, we show that it is actually possible to open this gap by considering the opposite limit of a {\em strongly} modulated magnetic field. In this case, unlike previous studies \cite{Snyman09,Guinea10,Lin11}, one cannot neglect the coupling between Dirac cones, which is directly responsible for this dramatic effect. As a consequence, the opening of the gap does not require the simultaneous presence of a scalar and a vector potential. To analyze this problem, we consider a vortex superlattice with fluxes $\pm \phi$ as depicted in Fig.~\ref{fig:lattice}. Our choice is motivated by the commensurability of the triangular and hexagonal structures and by the fact that this alternated pattern leads to the smallest possible unit cell of the superlattice. 
In the small-$\phi$  limit, we show that although the system remains gapless at first order, a gap proportional to $\phi^2$  may open, providing a nice example of a metal-insulator transition induced by a magnetic field in the honeycomb lattice. 
We derive the necessary and sufficient condition to open this gap in terms of the superlattice periodicity, and we give an expression of the gap at order two in the small-$\phi$ limit. 
When the size of the superlattice unit cell increases, {\it i.e.}, in the limit of vanishing vortex density $\nu$, we find that the gap vanishes as $\nu\ln \nu^{-1}$. 
%
%
%
%

Although obtained in a perturbative framework, our conclusions remain qualitatively valid for arbitrary fluxes as checked by exact diagonalizations. Furthermore, at $\phi=\phi_0/2$ ($\phi_0$ being the elementary flux quantum), the same gap-opening mechanism applies to the celebrated Kitaev model \cite{Kitaev06} studied in the context of topologically ordered systems.

The starting point of our study is the following tight-binding Hamiltonian
%
%
\begin{equation}
H=-\sum_{\langle i,j \rangle} t_{i,j} |i\rangle \langle j�|,
\label{fig:ham}
\end{equation}
%
%
where $|i\rangle$ denotes a spinless-electron state localized on site $i$. The sum is performed over all nearest-neighbor sites of the honeycomb lattice and the hopping term in the presence of a vector potential ${\bf A}$ is given by the so-called Peierls substitution \cite{Peierls33}: $t_{i,j}= t \, {\rm e}^{\frac{2{\rm i} \pi}{\phi_0} \int_i^j {\bf A}.{\rm d}{\bf l}}$. Thus, setting the flux and energy scales to unity 
($\phi_0=t=1$), the (oriented) product of the hopping terms over a closed loop is simply ${\rm e}^{2{\rm i} \pi \phi}$ where $\phi$ is the dimensionless magnetic flux inside the corresponding loop.

The vortex superlattice considered here is defined as follows. Let us assume that there is a flux $+\phi$ in the elementary plaquette centered in ${\bf r}$. Then, the superlattice ${\mathcal L}(p,q)$ is generated by requiring that the plaquette located at ${\bf r}+\bb_1/2$ contains a flux $-\phi$ and the one located at ${\bf r}+\bb_2$ contains a flux $+\phi$, where 
%
%
\begin{eqnarray}
\bb_1&=& 2(p \: \ba_1+q \: \ba_2), \\
\bb_2&=&-q \: \ba_1+(p+q) \: \ba_2.
\label{eq:unit_vectors}
\end{eqnarray}
%
%
Vectors $\ba_1$ and $\ba_2$ are primitive vectors of the honeycomb lattice (see Fig.~\ref{fig:lattice} for the case $p=q=1$) and $(p,q)$ are positive integers. In the following, without loss of generality, we  only consider the case $p\geqslant q$.
It is straightforward to check that the total flux per unit cell of ${\mathcal L}(p,q)$, spanned by $\bb_1$ and 
$\bb_2$, is zero. In addition, the vortex density defined as the number of vortices per unit cell is simply given by
%
%
\begin{equation}
\nu=\frac{1}{p^2+p q +q^2}.
\label{eq:nu}
\end{equation}
%
%
A convenient gauge choice realizing such a flux pattern can be obtained starting from an initial $+\phi$ plaquette center and by choosing  $t_{i,j}= {\rm e}^{2{\rm i} \pi \phi}$ for all links crossed by going $p$ times in direction $\ba_1$ and then $q$ times in the direction $\ba_2$. The orientation of the first link fixes all others since we wish to have a flux $-\phi$ in the final plaquette and zero in all intermediate ones. In other words, one creates a string of links carrying the flux which connects  a vortex to an antivortex.
As a side remark, let us note that with this gauge choice, one can study any value of the flux without changing the size of the unit cell, in deep contrast with the uniform field problem.  

As for any bipartite lattice, the spectrum of $H$ is symmetric with respect to the energy $\varepsilon=0$ for all $\phi$. For $\phi=0$, it consists of two symmetric bands \cite{Wallace47}
%
%
\begin{eqnarray}
\varepsilon_\pm (\bk)&=& \pm \Big\{3+2 \cos(\bk.\ba_1)+2 \cos(\bk.\ba_2) \nonumber \\
&&+2 \cos[\bk.(\ba_1-\ba_2)] \Big\}^{1/2}.
\label{eq:spectrum}
\end{eqnarray}
%
%
%
These symmetric bands touch at $\varepsilon=0$ when $\bk$ coincides with the so-called  Dirac points  
\mbox{$\bK=\frac{1}{3}\ba^*_1+\frac{2}{3}\ba^*_2$} and  \mbox{$\bK'=\frac{2}{3}\ba^*_1+\frac{1}{3}\ba^*_2$}, where $\ba^*_1$ and $\ba^*_2$ are primitive vectors of the reciprocal lattice associated to $\ba_1$ and $\ba_2$ \mbox{($\ba^*_i.\ba_j=2\pi \delta_{i,j}$)}. Consequently, the energy $\varepsilon=0$ is four-fold degenerate for $\phi=0$. Our goal is to determine the fate of these zero-energy states for $\phi \neq 0$.

To address this problem, we shall analyze perturbatively the small-$\phi$ limit. However, one can already predict that if the perturbation does not couple any of the two eigenstates corresponding to $\bK$ with the two eigenstates corresponding to $\bK'$,  the system will remain gapless {\em at all orders} for $\varepsilon=0$. Indeed,  in this case, the single-cone approximation proposed in Refs.~\cite{Snyman09,Guinea10,Lin11} can be made safely, leading to a finite gap only when a scalar as well as a vector potential are present. Thus, to open the gap, one must have a perturbing potential that couples these two twofold-degenerate subspaces. 
Since this potential has, by construction, the same periodicity as ${\mathcal L}(p,q)$, this  condition requires the existence of a reciprocal lattice vector associated to $\bb_1$ and $\bb_2$,  which equals $\bK' -\bK$.
It is then straightforward to show that this condition is strictly equivalent to
%
%
\begin{equation}
\frac{1}{\nu}=0 \mod 3,
\label{eq:condition}
\end{equation}
%
%
where $\nu$ is the vortex density defined in Eq.~(\ref{eq:nu}). Dirac states then have a momentum  \mbox{$\bk=0 \mod (\bb^*_1,\bb^*_2)$} where $\bb^*_1$ and $\bb^*_2$ are primitive vectors of the reciprocal lattice associated to $\bb_1$ and $\bb_2$ \mbox{($\bb_i^*.\bb_j=2\pi \delta_{i,j}$)}.
Let us underline that this is a necessary condition that might not be sufficient to open a gap but, as we shall  see, it is.

A naive first-order degenerate perturbation theory consists in considering the subspace spanned by the four Dirac states. There, one gets a finite gap \mbox{$\Delta(\bk=0)=2 \pi \phi \sqrt{\nu}$}. 
However, it is clear that condition (\ref{eq:condition}) together with the similar conic dispersions near $\bK$ and $\bK'$  implies that states in the vicinity of the Dirac cones are also coupled by the perturbation and one must look for $\bk \neq 0$ states that may have a lower gap. Of course, the corresponding subspace depends directly on the vector potential.  For the gauge choice described after Eq.~(\ref{eq:nu}), one finds that the state that has the lowest positive energy is found for $\bk_0= \frac{\phi}{2}\bb^*_2$. We checked by exact diagonalizations that this remarkable result is valid for any flux $\phi$ for the configurations $\mathcal{L}(p,q)$ with $\nu\leq 1/12$. The state with the lowest positive energy is therefore expected to always be found in this sector \cite{Doucot11}. Of course, the corresponding energy may be degenerate and may also be found for other momenta, as is the case for $\phi=1/2$.
%
%
\begin{table}[h]
\center
\begin{tabular}{c c c c}
\hline
\hline
$1/\nu$ & $p$ & $q$& $\Delta/(\pi \phi)^2$ \\ 
\hline
$3$ &   $1$  & $1$ & $1 /3$ \\  
$9$ &   $3$  & $0$ & $5/21$\\    
$12$ &   $2$  & $2$ & $1/6$\\    
$21$ & $4$  & $1$ &  $0.077586$ \\    
$27$ & $3$  & $3$ &  $0.061324$ \\    
$36$ & $6$  & $0$ & $11/130$\\    
\hline
\hline
\end{tabular}
\label{tab:mean_field_extremum}
\caption{Gap $\Delta$, at order $\phi^2$, for the first values of  $1/\nu$ satisfying Eq.~(\ref{eq:condition}). For $1/\nu=21$ and $27$, the gap cannot be expressed as a simple fraction and we only give the first digits obtained numerically.}
\end{table}
%
%

\begin{figure}[!t]
\includegraphics[width=0.8\columnwidth]{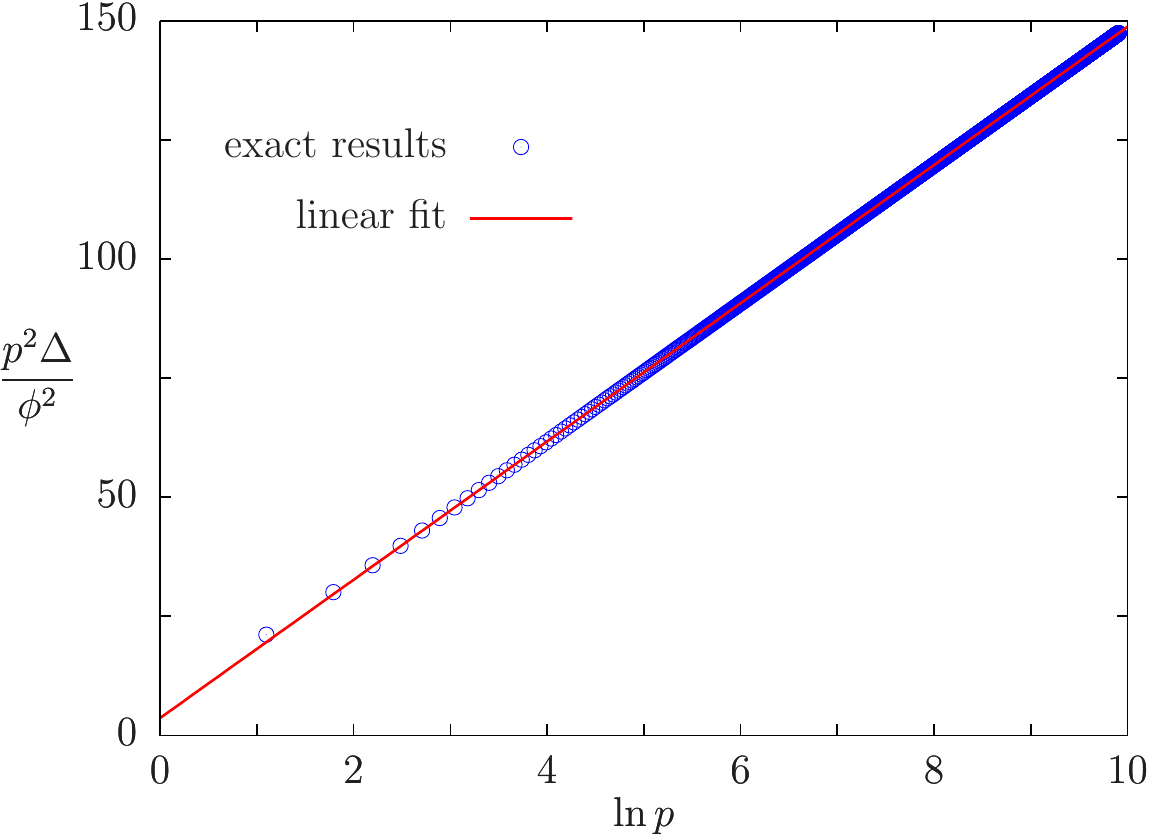}
\caption{(Color online) Behavior of $p^2 \Delta/\phi^2$ as a function of $\ln p$ (for $q=0$) in the small-$\phi$ limit. Exact results are obtained from Eq.~(\ref{eq:gap}) and the full line is a linear fit in good agreement with the conjecture discussed in the text.}
\label{fig:gapvsp}
\end{figure}

\begin{figure}[t]
\includegraphics[width=0.8\columnwidth]{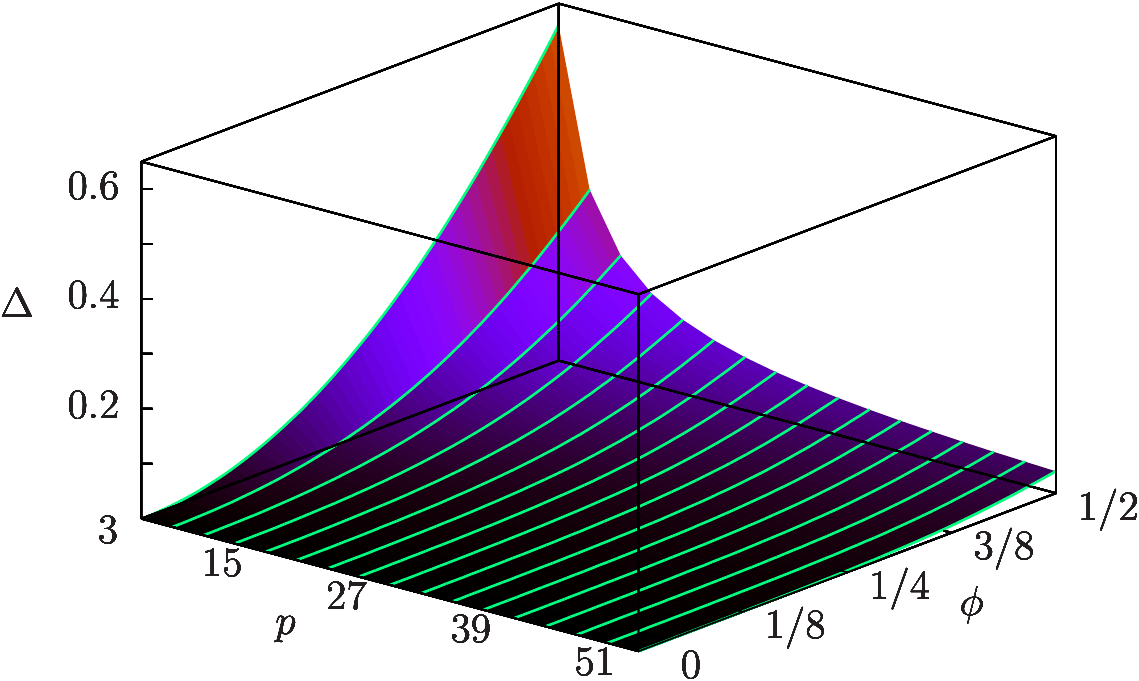}
\caption{(Color online) $\Delta$ as a function of $\phi$ and $p$ \mbox{(for $q=0$)}. The maximum is reached for $p=3$ and $\phi=1/2$  where \mbox{$\Delta \simeq 0.611132$}.
}
\label{fig:gap3D}
\end{figure}

At first order in $\phi$, one  gets $\Delta(\bk_0)=0$ so that the low-energy effect of the perturbation is simply to shift the Dirac cones \cite{Asano11} (without renormalizing the Fermi velocity at $\varepsilon=0$).
To go beyond, one has to consider the second-order degenerate perturbation theory in the \mbox{$\bk=\bk_0$} subspace.
Such an analysis involves the computation of matrix elements of the perturbation between all states belonging to this sector which, for arbitrary $p$ and $q$, is not an easy task. 
The expression of the gap for the first fillings satisfying  condition (\ref{eq:condition}) is given in Table I. 
Although in general it is difficult to get a simple expression of $\Delta$,  one can derive exact formulas for $q=0$ \mbox{($p$ being a multiple of 3)} that allow one to (numerically) investigate large unit cell systems that would be out of reach with exact diagonalizations. From now on, we will mainly focus on this subset of configurations for which the gap reads
%
%
\begin{equation}
\frac{\Delta(p,q=0)}{(\pi \phi)^2} =\mathcal{C}_p - \sqrt{\mathcal{B}_p^2 +\left(\mathcal{A}_p-\frac{1}{p}\right)^2},
\label{eq:gap}
\end{equation}
%
%
with 
%
%
\begin{widetext}
\begin{align}
\mathcal{A}_p & = \frac{1}{2 p^4} \sum\limits_{n=0}^{p-1} \sum\limits_{m=1}^{2p-1}
\frac{\xi(p,m)}{\varepsilon^2(m,n)} 
\Big\{
3 + 4\cos\big[\pi \big(\tfrac{m}{p} + \tfrac{2}{3}\big)\big]+2 \cos\big[2 \pi \big(\tfrac{m}{p} - \tfrac{1}{3}\big)\big]
\Big\}
\Big\{
1+\cos \big(\tfrac{2\pi n}{p} \big) + \cos\big[\tfrac{\pi}{p}(2n-m)\big]
\Big\},\\
\mathcal{B}_p & = \frac{1}{2 p^4} \sum\limits_{n=0}^{p-1} \sum\limits_{m=1}^{2p-1}
\frac{\xi(p,m)}{\varepsilon^2(m,n)} 
\Big\{
3 + 4\cos\big[\pi \big(\tfrac{m}{p} + \tfrac{2}{3}\big)\big]+2 \cos\big[2 \pi \big(\tfrac{m}{p} - \tfrac{1}{3}\big)\big]
\Big\}
\Big\{
\sin\big(\tfrac{2\pi n}{p} \big) + \sin\big[\tfrac{\pi}{p}(2n-m)\big]
\Big\},\\
\mathcal{C}_p & = \frac{4}{ p^4} \sum\limits_{n=0}^{p-1} \sum\limits_{m=1}^{2p-1}
\frac{\xi(p,m)}{\varepsilon^2(m,n)} 
\Big[
\cos\big(\tfrac{2\pi m}{p} \big) -\cos\big(\tfrac{\pi m}{p} \big)
\Big]
\Big\{
1+\cos \big(\tfrac{2\pi n}{p} \big) + \cos\big[\tfrac{\pi}{p}(2n-m)\big]
\Big\},
\end{align}

\end{widetext}
%
%
where the sum over $m$ is performed over odd integers only. For convenience, we also introduced \mbox{$\varepsilon^2(m,n)=\varepsilon^2_\pm ({\bf k}=\tfrac{m}{2p} \ba_1^*+\tfrac{n}{p} \ba_2^*)$} [see Eq.~(\ref{eq:spectrum})], and 
%
%
\begin{equation}
  \xi(p,m) = 
\begin{cases}
  \sin^{-2}\big(\tfrac{3\pi m}{2p}\big)  & \text{if } m \neq 0 \mod \frac{p}{3} \text{,}\\
  \frac{1-(-1)^{p/3}}{2} & \text{otherwise.}
\end{cases}
\end{equation}
%
%
In the large-$p$ limit,  it is clear that $\Delta$ vanishes since one has to recover the spectral properties of the zero-flux problem. To analyze this infinitely diluted vortex limit, we computed the gap using Eq.~(\ref{eq:gap}) up to \mbox{$p= 20000$}. 
A close inspection of $\mathcal{A}_p$, $\mathcal{B}_p$, and $\mathcal{C}_p$ led us to conjecture that the gap vanishes as \mbox{$\Delta/\phi^2 \sim \nu \ln \nu^{-1}$} in the large-$p=1/\sqrt{\nu}$ limit. A convincing check of this result is, however, displayed in Fig.~\ref{fig:gapvsp}.
A natural question that arises at this stage concerns the behavior of the gap away from the perturbative regime analyzed up to now. To investigate arbitrary fluxes, one must diagonalize $H$ numerically but the main advantage is that one only has to consider the subspace corresponding to $\bk=\bk_0$ \cite{Doucot11} where the lowest-positive energy state lies. However, for arbitrary fluxes, one is restricted to small values of $p$ since the number of sites per unit cell is $4 p^2$ and we need the full spectrum of the $\bk=\bk_0$ subspace. 
%
%
%
%
In Fig.~\ref{fig:gap3D}, we display the behavior of the gap as a function of $\phi$ and for $q=0$ and $p=3,6,\dots,51$. As can be seen, $\Delta$ is a monotonously decreasing (increasing) function of $p$ (of $\phi$ in the interval $[0,1/2]$). We have also observed that the way $\Delta$ vani\-shes when $p$ increases depends on $\phi$. However, the lack of large-$p$ data prevents to perform a sound analysis of these behaviors.

As already observed in the small-$\phi$ limit (see Table I), the maximum value of the gap is obtained for the largest vortex density satisfying Eq.~(\ref{eq:condition}), {\it i.e.}, $\nu=1/3$, but it is also obtained for the largest possible flux, {\it i.e.}, $\phi=1/2$. Denoting $x^*$, the smallest positive root of the following polynomial
%
%
\begin{equation}
P(x)=x^6-18 \: x^5+117 \: x^4-340 \: x^3+428 \: x^2-176 \: x+16,
\end{equation}
%
%
one gets 
%
%
\begin{equation}
\Delta(p=1,q=1,\phi=1/2)=2 \sqrt{x^*}\simeq 0.70884.
\end{equation}
%
%
%
%
%
%
Surprisingly enough, the problem considered here for $\phi=1/2$ is directly connected to the  spin $1/2$ model introduced by Kitaev in 2006 \cite{Kitaev06}. Using a Majorana fermionic representation of Pauli matrices, the Kitaev model can be mapped onto a free-fermion model in a $\mathbb{Z}_2$ gauge field. As a consequence, the value of the (effective) flux in each elementary plaquette is restricted to $\phi=0$ or $1/2$. This correspondence allowed Kitaev to identify the vortex configuration where the ground state of his system (Fermi sea at half-filling in the present electron language) lies. Indeed, as suggested early on in the flux-phase framework \cite{Hasegawa89,Montambaux89,Hasegawa90}, the lowest energy at half-filling is obtained for $\phi=0$. 
In this problem, the more general question was as follows: for a given electron density, what is the flux density (and the flux pattern) that minimizes the energy~?  Although the answer has been provided by Lieb \cite{Lieb94} for the special case of half-filling, exact results are still missing for arbitrary electron density.

The present study raises a complementary question: given a flux density, what is the flux pattern that maximizes the gap~? Undoubtedly, this question is even more difficult and the answer likely depends on the electron density. For the Kitaev model (half-filling and $\phi=0,1/2$), we investigated several periodic configurations corresponding to fixed flux density $\nu$ satisfying (\ref{eq:condition}), and we are led to conjecture that the flux  pattern maximizing the gap is always ${\mathcal L}(p,q)$. Note that this flux pattern also minimizes the energy. One way to understand this result is to argue that the vortex-vortex interaction for $\phi=1/2$ is repulsive so that it seems natural to find a triangular (Abrikosov-like) superlattice as an optimal pattern. 
However, it would be valuable to prove this result rigorously as well as  finding gapped flux configurations for arbitrary $\nu$.
It would also be worth adding further hopping processes as discussed in Ref.~\cite{Haldane88} that may give rise to a nontrivial insulator.
Such considerations are clearly beyond the scope of the present work but we hope to have underlined that interesting phenomena may occur for nontrivial vortex configurations in the honeycomb lattice (see also Ref.~\cite{Lahtinen11} for related studies of the Kitaev model). An obvious consequence of our results for the Kitaev model is that there must be a finite gapped region around the point where a gap is induced by the vortex superlattice ${\mathcal L}(p,q)$. This is due to the fact that an insulator, as the one considered here, is robust to small deformations (for example, anisotropies in the hopping elements).

One must also wonder how to observe this metal-insulator transition induced by a vortex superlattice in the honeycomb lattice. Obviously, the main difficulty is the realization of the  superlattice with the {\it ad hoc}  parameters. The most realistic choice would be a flux $\phi=1/2$ for which vortices and antivortices are equivalent, so that one can use the exact correspondence to the Kitaev model for which many experimental proposals exist \cite{Duan03,Micheli06,Jackeli09,You10}. Otherwise, in the context of graphene, a type-II superconductor might be used in the mixed state where the Abrikosov vortex lattice is found. Then, given $\phi=1/2$, one could think about gluing a graphene sheet on top of the superconductor. However, one faces the problem that the vortex core is much larger than a single elementary plaquette. Therefore, one has to investigate the gap-opening problem in the presence of extended though localized flux spots.
Another appealing approach would be to consider optical flux lattices recently suggested in Refs.~\cite{Cooper11_1,Cooper11_2} that seem especially adapted to our problem. 

Finally, given the occurrence of Dirac points in many experimental devices (see Ref.~\cite{Asano11} for a recent discussion), we hope that the present work will motivate further investigations concerning the fate of these singularities in the presence of a vortex superlattice.\\
 
\acknowledgments

We would like to thank B. Dou{\c c}ot and graphene's theory group of Laboratoire de Physique des Solides (Universit\'e Paris-Sud) for stimulating discussions. K.~P. Schmidt and M. Kamfor acknowledge financial support from the DFG and thank ESF and EuroHorcs for funding through the EURYI.


\begin{thebibliography}{24}%
\makeatletter
\providecommand \@ifxundefined [1]{%
 \@ifx{#1\undefined}
}%
\providecommand \@ifnum [1]{%
 \ifnum #1\expandafter \@firstoftwo
 \else \expandafter \@secondoftwo
 \fi
}%
\providecommand \@ifx [1]{%
 \ifx #1\expandafter \@firstoftwo
 \else \expandafter \@secondoftwo
 \fi
}%
\providecommand \natexlab [1]{#1}%
\providecommand \enquote  [1]{``#1''}%
\providecommand \bibnamefont  [1]{#1}%
\providecommand \bibfnamefont [1]{#1}%
\providecommand \citenamefont [1]{#1}%
\providecommand \href@noop [0]{\@secondoftwo}%
\providecommand \href [0]{\begingroup \@sanitize@url \@href}%
\providecommand \@href[1]{\@@startlink{#1}\@@href}%
\providecommand \@@href[1]{\endgroup#1\@@endlink}%
\providecommand \@sanitize@url [0]{\catcode `\\12\catcode `\$12\catcode
  `\&12\catcode `\#12\catcode `\^12\catcode `\_12\catcode `\%12\relax}%
\providecommand \@@startlink[1]{}%
\providecommand \@@endlink[0]{}%
\providecommand \url  [0]{\begingroup\@sanitize@url \@url }%
\providecommand \@url [1]{\endgroup\@href {#1}{\urlprefix }}%
\providecommand \urlprefix  [0]{URL }%
\providecommand \Eprint [0]{\href }%
\providecommand \doibase [0]{http://dx.doi.org/}%
\providecommand \selectlanguage [0]{\@gobble}%
\providecommand \bibinfo  [0]{\@secondoftwo}%
\providecommand \bibfield  [0]{\@secondoftwo}%
\providecommand \translation [1]{[#1]}%
\providecommand \BibitemOpen [0]{}%
\providecommand \bibitemStop [0]{}%
\providecommand \bibitemNoStop [0]{.\EOS\space}%
\providecommand \EOS [0]{\spacefactor3000\relax}%
\providecommand \BibitemShut  [1]{\csname bibitem#1\endcsname}%
\let\auto@bib@innerbib\@empty
\bibitem [{\citenamefont {{K. S. Novoselov {\it et al.}}}(2004)}]{Novoselov04}%
  \BibitemOpen
  \bibfield  {author} {\bibinfo {author} {\bibnamefont {{K. S. Novoselov {\it
  et al.}}}},\ }\href {\doibase 10.1126/science.1102896} {\bibfield  {journal}
  {\bibinfo  {journal} {Science}\ }\textbf {\bibinfo {volume} {306}},\ \bibinfo
  {pages} {666} (\bibinfo {year} {2004})}\BibitemShut {NoStop}%
\bibitem [{\citenamefont {{Castro Neto}}\ \emph {et~al.}(2009)\citenamefont
  {{Castro Neto}}, \citenamefont {Guinea}, \citenamefont {Peres}, \citenamefont
  {Novoselov},\ and\ \citenamefont {Geim}}]{CastroNeto09}%
  \BibitemOpen
  \bibfield  {author} {\bibinfo {author} {\bibfnamefont {A.~H.}\ \bibnamefont
  {{Castro Neto}}}, \bibinfo {author} {\bibfnamefont {F.}~\bibnamefont
  {Guinea}}, \bibinfo {author} {\bibfnamefont {N.~M.~R.}\ \bibnamefont
  {Peres}}, \bibinfo {author} {\bibfnamefont {K.~S.}\ \bibnamefont
  {Novoselov}}, \ and\ \bibinfo {author} {\bibfnamefont {A.~K.}\ \bibnamefont
  {Geim}},\ }\href {\doibase 10.1103/RevModPhys.81.109} {\bibfield  {journal}
  {\bibinfo  {journal} {Rev. Mod. Phys.}\ }\textbf {\bibinfo {volume} {81}},\
  \bibinfo {pages} {109} (\bibinfo {year} {2009})}\BibitemShut {NoStop}%
\bibitem [{\citenamefont {Wallace}(1947)}]{Wallace47}%
  \BibitemOpen
  \bibfield  {author} {\bibinfo {author} {\bibfnamefont {P.~R.}\ \bibnamefont
  {Wallace}},\ }\href {\doibase 10.1103/PhysRev.71.622} {\bibfield  {journal}
  {\bibinfo  {journal} {Phys. Rev.}\ }\textbf {\bibinfo {volume} {71}},\
  \bibinfo {pages} {622} (\bibinfo {year} {1947})}\BibitemShut {NoStop}%
\bibitem [{\citenamefont {Rammal}(1985)}]{Rammal85}%
  \BibitemOpen
  \bibfield  {author} {\bibinfo {author} {\bibfnamefont {R.}~\bibnamefont
  {Rammal}},\ }\href {\doibase 10.1051/jphys:019850046080134500} {\bibfield
  {journal} {\bibinfo  {journal} {J. Phys. France}\ }\textbf {\bibinfo {volume}
  {46}},\ \bibinfo {pages} {1345} (\bibinfo {year} {1985})}\BibitemShut
  {NoStop}%
\bibitem [{\citenamefont {Esaki}\ \emph {et~al.}(2009)\citenamefont {Esaki},
  \citenamefont {Sato}, \citenamefont {Kohmoto},\ and\ \citenamefont
  {Halperin}}]{Esaki09}%
  \BibitemOpen
  \bibfield  {author} {\bibinfo {author} {\bibfnamefont {K.}~\bibnamefont
  {Esaki}}, \bibinfo {author} {\bibfnamefont {M.}~\bibnamefont {Sato}},
  \bibinfo {author} {\bibfnamefont {M.}~\bibnamefont {Kohmoto}}, \ and\
  \bibinfo {author} {\bibfnamefont {B.~I.}\ \bibnamefont {Halperin}},\ }\href
  {\doibase 10.1103/PhysRevB.80.125405} {\bibfield  {journal} {\bibinfo
  {journal} {Phys. Rev. B}\ }\textbf {\bibinfo {volume} {80}},\ \bibinfo
  {pages} {125405} (\bibinfo {year} {2009})}\BibitemShut {NoStop}%
\bibitem [{\citenamefont {Snyman}(2009)}]{Snyman09}%
  \BibitemOpen
  \bibfield  {author} {\bibinfo {author} {\bibfnamefont {I.}~\bibnamefont
  {Snyman}},\ }\href {\doibase 10.1103/PhysRevB.80.054303} {\bibfield
  {journal} {\bibinfo  {journal} {Phys. Rev. B}\ }\textbf {\bibinfo {volume}
  {80}},\ \bibinfo {pages} {054303} (\bibinfo {year} {2009})}\BibitemShut
  {NoStop}%
\bibitem [{\citenamefont {Guinea}\ and\ \citenamefont {Low}(2010)}]{Guinea10}%
  \BibitemOpen
  \bibfield  {author} {\bibinfo {author} {\bibfnamefont {F.}~\bibnamefont
  {Guinea}}\ and\ \bibinfo {author} {\bibfnamefont {T.}~\bibnamefont {Low}},\
  }\href {\doibase 10.1098/rsta.2010.0214} {\bibfield  {journal} {\bibinfo
  {journal} {Phil. Trans. R. Soc. A}\ }\textbf {\bibinfo {volume} {368}},\
  \bibinfo {pages} {5391} (\bibinfo {year} {2010})}\BibitemShut {NoStop}%
\bibitem [{\citenamefont {Lin}\ \emph {et~al.}()\citenamefont {Lin},
  \citenamefont {Wang}, \citenamefont {Pan},\ and\ \citenamefont {Xu}}]{Lin11}%
  \BibitemOpen
  \bibfield  {author} {\bibinfo {author} {\bibfnamefont {X.}~\bibnamefont
  {Lin}}, \bibinfo {author} {\bibfnamefont {H.}~\bibnamefont {Wang}}, \bibinfo
  {author} {\bibfnamefont {H.}~\bibnamefont {Pan}}, \ and\ \bibinfo {author}
  {\bibfnamefont {H.}~\bibnamefont {Xu}},\ }\href@noop {} {}\bibinfo {note}
  {\href{http://arxiv.org/abs/1106.0083}{arXiv:1106.0083}}\BibitemShut
  {NoStop}%
\bibitem [{\citenamefont {Kitaev}(2006)}]{Kitaev06}%
  \BibitemOpen
  \bibfield  {author} {\bibinfo {author} {\bibfnamefont {A.}~\bibnamefont
  {Kitaev}},\ }\href {\doibase 10.1016/j.aop.2005.10.005} {\bibfield  {journal}
  {\bibinfo  {journal} {Ann. Phys. (N.Y.)}\ }\textbf {\bibinfo {volume}
  {321}},\ \bibinfo {pages} {2} (\bibinfo {year} {2006})}\BibitemShut {NoStop}%
\bibitem [{\citenamefont {Peierls}(1933)}]{Peierls33}%
  \BibitemOpen
  \bibfield  {author} {\bibinfo {author} {\bibfnamefont {R.~E.}\ \bibnamefont
  {Peierls}},\ }\href {\doibase 10.1007/BF01342591} {\bibfield  {journal}
  {\bibinfo  {journal} {Z. Phys.}\ }\textbf {\bibinfo {volume} {80}},\ \bibinfo
  {pages} {763} (\bibinfo {year} {1933})}\BibitemShut {NoStop}%
\bibitem [{\citenamefont {{B. Dou{\c c}ot {\it et al.}}}()}]{Doucot11}%
  \BibitemOpen
  \bibfield  {author} {\bibinfo {author} {\bibnamefont {{B. Dou{\c c}ot {\it et
  al.}}}},\ }\href@noop {} {}\bibinfo {note} {(unpublished)}\BibitemShut
  {NoStop}%
\bibitem [{\citenamefont {Asano}\ and\ \citenamefont {Hotta}(2011)}]{Asano11}%
  \BibitemOpen
  \bibfield  {author} {\bibinfo {author} {\bibfnamefont {K.}~\bibnamefont
  {Asano}}\ and\ \bibinfo {author} {\bibfnamefont {C.}~\bibnamefont {Hotta}},\
  }\href {\doibase 10.1103/PhysRevB.83.245125} {\bibfield  {journal} {\bibinfo
  {journal} {Phys. Rev. B}\ }\textbf {\bibinfo {volume} {83}},\ \bibinfo
  {pages} {245125} (\bibinfo {year} {2011})}\BibitemShut {NoStop}%
\bibitem [{\citenamefont {Hasegawa}\ \emph {et~al.}(1989)\citenamefont
  {Hasegawa}, \citenamefont {Lederer}, \citenamefont {Rice},\ and\
  \citenamefont {Wiegmann}}]{Hasegawa89}%
  \BibitemOpen
  \bibfield  {author} {\bibinfo {author} {\bibfnamefont {Y.}~\bibnamefont
  {Hasegawa}}, \bibinfo {author} {\bibfnamefont {P.}~\bibnamefont {Lederer}},
  \bibinfo {author} {\bibfnamefont {T.~M.}\ \bibnamefont {Rice}}, \ and\
  \bibinfo {author} {\bibfnamefont {P.~B.}\ \bibnamefont {Wiegmann}},\ }\href
  {\doibase 10.1103/PhysRevLett.63.907} {\bibfield  {journal} {\bibinfo
  {journal} {Phys. Rev. Lett.}\ }\textbf {\bibinfo {volume} {63}},\ \bibinfo
  {pages} {907} (\bibinfo {year} {1989})}\BibitemShut {NoStop}%
\bibitem [{\citenamefont {Montambaux}(1989)}]{Montambaux89}%
  \BibitemOpen
  \bibfield  {author} {\bibinfo {author} {\bibfnamefont {G.}~\bibnamefont
  {Montambaux}},\ }\href {\doibase 10.1103/PhysRevLett.63.1657} {\bibfield
  {journal} {\bibinfo  {journal} {Phys. Rev. Lett.}\ }\textbf {\bibinfo
  {volume} {63}},\ \bibinfo {pages} {1657} (\bibinfo {year}
  {1989})}\BibitemShut {NoStop}%
\bibitem [{\citenamefont {Hasegawa}\ \emph {et~al.}(1990)\citenamefont
  {Hasegawa}, \citenamefont {Hatsugai}, \citenamefont {Kohmoto},\ and\
  \citenamefont {Montambaux}}]{Hasegawa90}%
  \BibitemOpen
  \bibfield  {author} {\bibinfo {author} {\bibfnamefont {Y.}~\bibnamefont
  {Hasegawa}}, \bibinfo {author} {\bibfnamefont {Y.}~\bibnamefont {Hatsugai}},
  \bibinfo {author} {\bibfnamefont {M.}~\bibnamefont {Kohmoto}}, \ and\
  \bibinfo {author} {\bibfnamefont {G.}~\bibnamefont {Montambaux}},\ }\href
  {\doibase 10.1103/PhysRevB.41.9174} {\bibfield  {journal} {\bibinfo
  {journal} {Phys. Rev. B}\ }\textbf {\bibinfo {volume} {41}},\ \bibinfo
  {pages} {9174} (\bibinfo {year} {1990})}\BibitemShut {NoStop}%
\bibitem [{\citenamefont {Lieb}(1994)}]{Lieb94}%
  \BibitemOpen
  \bibfield  {author} {\bibinfo {author} {\bibfnamefont {E.~H.}\ \bibnamefont
  {Lieb}},\ }\href {\doibase 10.1103/PhysRevLett.73.2158} {\bibfield  {journal}
  {\bibinfo  {journal} {Phys. Rev. Lett.}\ }\textbf {\bibinfo {volume} {73}},\
  \bibinfo {pages} {2158} (\bibinfo {year} {1994})}\BibitemShut {NoStop}%
\bibitem [{\citenamefont {Haldane}(1988)}]{Haldane88}%
  \BibitemOpen
  \bibfield  {author} {\bibinfo {author} {\bibfnamefont {F.~D.~M.}\
  \bibnamefont {Haldane}},\ }\href {\doibase 10.1103/PhysRevLett.61.2015}
  {\bibfield  {journal} {\bibinfo  {journal} {Phys. Rev. Lett.}\ }\textbf
  {\bibinfo {volume} {61}},\ \bibinfo {pages} {2015} (\bibinfo {year}
  {1988})}\BibitemShut {NoStop}%
\bibitem [{\citenamefont {{V. Lahtinen {\it et al.}}}()}]{Lahtinen11}%
  \BibitemOpen
  \bibfield  {author} {\bibinfo {author} {\bibnamefont {{V. Lahtinen {\it et
  al.}}}},\ }\href@noop {} {}\bibinfo {note} {(unpublished)}\BibitemShut
  {NoStop}%
\bibitem [{\citenamefont {Duan}\ \emph {et~al.}(2003)\citenamefont {Duan},
  \citenamefont {Demler},\ and\ \citenamefont {Lukin}}]{Duan03}%
  \BibitemOpen
  \bibfield  {author} {\bibinfo {author} {\bibfnamefont {L.-M.}\ \bibnamefont
  {Duan}}, \bibinfo {author} {\bibfnamefont {E.}~\bibnamefont {Demler}}, \ and\
  \bibinfo {author} {\bibfnamefont {M.~D.}\ \bibnamefont {Lukin}},\ }\href
  {\doibase 10.1103/PhysRevLett.91.090402} {\bibfield  {journal} {\bibinfo
  {journal} {Phys. Rev. Lett.}\ }\textbf {\bibinfo {volume} {91}},\ \bibinfo
  {pages} {090402} (\bibinfo {year} {2003})}\BibitemShut {NoStop}%
\bibitem [{\citenamefont {Micheli}\ \emph {et~al.}(2006)\citenamefont
  {Micheli}, \citenamefont {Brennen},\ and\ \citenamefont
  {Zoller}}]{Micheli06}%
  \BibitemOpen
  \bibfield  {author} {\bibinfo {author} {\bibfnamefont {A.}~\bibnamefont
  {Micheli}}, \bibinfo {author} {\bibfnamefont {G.~K.}\ \bibnamefont
  {Brennen}}, \ and\ \bibinfo {author} {\bibfnamefont {P.}~\bibnamefont
  {Zoller}},\ }\href {\doibase 10.1038/nphys287} {\bibfield  {journal}
  {\bibinfo  {journal} {Nat. Phys.}\ }\textbf {\bibinfo {volume} {2}},\
  \bibinfo {pages} {341} (\bibinfo {year} {2006})}\BibitemShut {NoStop}%
\bibitem [{\citenamefont {Jackeli}\ and\ \citenamefont
  {Khaliullin}(2009)}]{Jackeli09}%
  \BibitemOpen
  \bibfield  {author} {\bibinfo {author} {\bibfnamefont {G.}~\bibnamefont
  {Jackeli}}\ and\ \bibinfo {author} {\bibfnamefont {G.}~\bibnamefont
  {Khaliullin}},\ }\href {\doibase 10.1103/PhysRevLett.105.027204} {\bibfield
  {journal} {\bibinfo  {journal} {Phys. Rev. Lett.}\ }\textbf {\bibinfo
  {volume} {102}},\ \bibinfo {pages} {017205} (\bibinfo {year}
  {2009})}\BibitemShut {NoStop}%
\bibitem [{\citenamefont {You}\ \emph {et~al.}(2010)\citenamefont {You},
  \citenamefont {Shi}, \citenamefont {Hu},\ and\ \citenamefont {Nori}}]{You10}%
  \BibitemOpen
  \bibfield  {author} {\bibinfo {author} {\bibfnamefont {J.~Q.}\ \bibnamefont
  {You}}, \bibinfo {author} {\bibfnamefont {X.-F.}\ \bibnamefont {Shi}},
  \bibinfo {author} {\bibfnamefont {X.}~\bibnamefont {Hu}}, \ and\ \bibinfo
  {author} {\bibfnamefont {F.}~\bibnamefont {Nori}},\ }\href {\doibase
  10.1103/PhysRevB.81.014505} {\bibfield  {journal} {\bibinfo  {journal} {Phys.
  Rev. B}\ }\textbf {\bibinfo {volume} {81}},\ \bibinfo {pages} {014505}
  (\bibinfo {year} {2010})}\BibitemShut {NoStop}%
\bibitem [{\citenamefont {Cooper}(2011)}]{Cooper11_1}%
  \BibitemOpen
  \bibfield  {author} {\bibinfo {author} {\bibfnamefont {N.~R.}\ \bibnamefont
  {Cooper}},\ }\href {\doibase 10.1103/PhysRevLett.106.175301} {\bibfield
  {journal} {\bibinfo  {journal} {Phys. Rev. Lett.}\ }\textbf {\bibinfo
  {volume} {106}},\ \bibinfo {pages} {175301} (\bibinfo {year}
  {2011})}\BibitemShut {NoStop}%
\bibitem [{\citenamefont {Cooper}\ and\ \citenamefont
  {Dalibard}(2011)}]{Cooper11_2}%
  \BibitemOpen
  \bibfield  {author} {\bibinfo {author} {\bibfnamefont {N.~R.}\ \bibnamefont
  {Cooper}}\ and\ \bibinfo {author} {\bibfnamefont {J.}~\bibnamefont
  {Dalibard}},\ }\href {\doibase 10.1209/0295-5075/95/66004} {\bibfield
  {journal} {\bibinfo  {journal} {Europhys. Lett.}\ }\textbf {\bibinfo {volume}
  {95}},\ \bibinfo {pages} {66004} (\bibinfo {year} {2011})}\BibitemShut
  {NoStop}%
\end{thebibliography}

%

\end{document}